\documentclass[final,5p,times,twocolumn,authoryear]{elsarticle}

\usepackage{graphicx}
\usepackage{natbib}
\usepackage[colorlinks=true]{hyperref}
\usepackage{txfonts}
\usepackage{subfigure}
\usepackage{float}
\usepackage{bigints}
\usepackage{twoopt}
\usepackage{xspace}
\usepackage{balance}
\usepackage{upgreek}
\usepackage{multirow}
\usepackage{lineno}
\usepackage[procnames]{listings}
\usepackage{amssymb}
\usepackage{amsmath}
\usepackage[svgnames]{xcolor}
\usepackage{soul}
\usepackage{siunitx}
  
\journal{Astronomy and Computing}

\definecolor{keywords}{RGB}{255,0,90}
\definecolor{listinggray}{gray}{0.9}
\definecolor{lbcolor}{rgb}{0.9,0.9,0.9}

\definecolor{redUniBo}{RGB}{187, 46, 41}

\lstdefinestyle{C++} { language=C++, backgroundcolor=\color{lbcolor},
  basicstyle=\scriptsize\upshape\ttfamily, commentstyle=\color{blue},
  classoffset=1, morekeywords={cosmobl, DirCosmo, DirLoc, Cosmology,
    par, Object, Galaxy, Catalogue, random_catalogue_box,
    TwoPointCorrelation}, keywordstyle=\color{ForestGreen},
  classoffset=2, morekeywords={fsigma8, nObjects, setParameters,
    measure_xi, write_xi}, keywordstyle=\color{red}, classoffset=0,
  tabsize=2, captionpos=b, frame=lines, frameround=fttt, numbers=left,
  numberstyle=\tiny, numbersep=5pt, breaklines=true,
  showstringspaces=false }

\lstdefinestyle{Python} { language=Python,
  backgroundcolor=\color{lbcolor},
  basicstyle=\scriptsize\upshape\ttfamily, commentstyle=\color{blue},
  classoffset=1, morekeywords={Cosmology, pyCosmologyCBL},
  keywordstyle=\color{ForestGreen}, classoffset=2, morekeywords={D_C},
  keywordstyle=\color{red}, classoffset=0, tabsize=2, captionpos=b,
  frame=lines, numbers=left, numberstyle=\tiny, numbersep=5pt,
  breaklines=true, showstringspaces=false, procnamekeys={def,class} }

\begin{document}

\lstset{language=C++}

\begin{frontmatter}

  \title{Artificial Neural Networks for Galaxy Clustering:\\ Learning
    from the two-point correlation function of BOSS galaxies}
  
  \author[add1]{Niccol\'{o} Veronesi}
  \ead{veronesi@strw.leidenuniv.nl} \author[add2,add3,add4]{Federico
    Marulli} \author[add5,add6]{Alfondo Veropalumbo}
  \author[add2,add3,add4]{Lauro Moscardini} \address[add1]{Leiden
    Observatory, Leiden University, PO Box 9513, 2300 RA Leiden, The
    Netherlands} \address[add2]{Dipartimento di Fisica e Astronomia
    ``Augusto Righi'' - Alma Mater Studiorum Universit\`{a} di
    Bologna, via Piero Gobetti 93/2, I-40129 Bologna, Italy}
  \address[add3]{INAF - Osservatorio di Astrofisica e Scienza dello
    Spazio di Bologna, via Piero Gobetti 93/3, I-40129 Bologna, Italy}
  \address[add4]{INFN - Sezione di Bologna, viale Berti Pichat 6/2,
    I-40127 Bologna, Italy} \address[add5]{Dipartimento di Fisica,
    Universit\`{a} degli Studi Roma Tre, via della Vasca Navale 84,
    I-00146 Rome, Italy} \address[add6]{INFN - Sezione di Roma Tre,
    via della Vasca Navale 84, I-00146 Rome, Italy}


  \begin{abstract}
{The increasingly large amount of cosmological data coming from
  ground-based and space-borne telescopes requires highly efficient
  and fast enough data analysis techniques to maximise the scientific
  exploitation.}
{In this work, we explore the capabilities of supervised machine
  learning algorithms to learn the properties of the large-scale
  structure of the Universe, aiming at constraining the matter density
  parameter, $\Omega_m$.}
{We implement a new Artificial Neural Network for a regression data
  analysis, and train it on a large set of galaxy two-point
  correlation functions in standard cosmologies with different values
  of $\Omega_m$. The training set is constructed from log-normal mock
  catalogues which reproduce the clustering of the Baryon Oscillation
  Spectroscopic Survey (BOSS) galaxies. The presented statistical
  method requires no specific analytical model to construct the
  likelihood function, and runs with negligible computational cost,
  after training.}
{We test this new Artificial Neural Network on real BOSS data, finding
  $\Omega_m=0.309\pm0.008$, which is  consistent with standard
  analysis results. }
  \end{abstract}
  

\end{frontmatter}


\section{Introduction}
\label{sec:intro}

One of the biggest challenges of modern cosmology is to accurately
estimate the standard cosmological model parameters and, possibly, to
discriminate among alternative cosmological frameworks. Fast and
accurate statistical methods are required to maximise the scientific
exploitation of the cosmological probes of the large-scale structure
of the Universe. During the last decades, increasingly large surveys
have been conducted both with ground-based and space-borne telescopes,
and a huge amount of data is expected from next-generation projects,
like e.g. Euclid \citep{Laureijs2011, blanchard2020} and the Vera
C. Rubin Observatory \citep{lsst2021}.  This scenario, in which the
amount of available data is expected to keep growing with exponential
rate, suggests that machine learning techniques shall play a key role
in cosmological data analysis, due to the fact that their reliability
and precision strongly depend on the quantity and variety of inputs
they are given.

According to the standard cosmological scenario, the evolution of
density perturbations started from an almost Gaussian distribution,
primarily described by its variance. In configuration space, the
variance of the field is the two-point correlation function (2PCF),
which is a function of the magnitude of the comoving distance between
objects, $r$. At large enough scales, the matter distribution can
still be approximated as Gaussian, and thus the 2PCF contains most of
the information of the density field.

The 2PCF and its analogous in Fourier space, the power spectrum, have
been the focus of several cosmological analyses of observed catalogues
of extra-galactic sources \citep[see e.g.][and references
  therein]{Totsuji1969, Peebles1974, Hawkins2003, Parkinson2012,
  Bel2014, Alam2016, Pezzotta2017, Mohammad2018, Gil2020,
  Marulli2021}.  The standard way to infer constraints on cosmological
parameters from the measured 2PCF is by comparison with a physically
motivated model through an analytical likelihood function. The latter
should account for all possible observational effects, including both
statistical and systematic uncertainties. In this work, we investigate
an alternative data analysis method, based on a supervised machine
learning technique, which does not require a customized likelihood
function to model the 2PCF. Specifically, the 2PCF shape will be
modelled with an Artificial Neural Network (NN), trained on a
sufficiently large set of measurements from mock data sets generated
at different cosmologies.

  Supervised machine learning methods can be
  considered as a complementary approach to standard data analysis
  techniques for the investigation of the large-scale structure of the
  Universe. There are two main issues in applying machine learning
  algorithms in this context. Firstly, the simulated galaxy maps used
  to train the NNs have to be sufficiently accurate at all scales of
  interest, including all the relevant observational effects
  characterizing the surveys to be analysed. Indeed, the NN outputs
  rely only on the {\em examples} provided for the training phase, and
  not according to any specific instructions \citep{Samuel1959,
    Goodfellow2016}. The accuracy of the output depends on the level
  of reliability of the training set, while the precision increases
  when the amount of training data set increases. Moreover,
  significant computational resources are required to construct the
  mock data sets in a high enough number of test cosmologies.

  On the other hand, both the construction of the
  mock data sets and the NN training and validation have to be done
  only once. After that, the NN can produce outputs with negligible
  computational cost, and can exploit all physical scales on which the
  mock data sets are reliable, that is possibly beyond the domain in
  which an analytic likelihood is available. These represent the key
  advantages of this novel, complementary data analysis technique.

Machine learning based methods should provide an effective tool in
cosmological investigations based on increasingly large amounts of
cosmological data from extra-galactic surveys \citep[e.g.][]{Cole2005,
  Parkinson2012, Anderson2014, Kern2017,Ntampaka2019,Ishida2019,
  Hassan2020, Villaescusa2021}.  In fact, these techniques have
already been exploited for the analysis of the large-scale structure
of the Universe \citep[see e.g.][]{Aragon2019, Tsizh2020}. In some
cases, machine learning models have been trained and tested on
simulated mock catalogues to obtain as output the cosmological
parameters those simulations had been constructed with
\citep[e.g.][]{Ravanbakhsh2018, Pan2020}. These works demonstrated
that the machine learning approach is powerful enough to infer tight
constraints on cosmological model parameters, when the training set is
the distribution of matter in a three-dimensional grid.

The method we are presenting in this work is different in this
respect, as our input training set consists of 2PCF measurements
estimated from mock galaxy catalogues. The rationale of this choice is
to help the network to efficiently learn the mapping between the
cosmological model and the corresponding galaxy catalogue by
exploiting the information compression provided by the second-order
statistics of the density field. The goal of this
  work is to implement, train, validate and test a new NN of this
  kind, and to investigate its capability in providing constraints on
  the matter density parameter, $\Omega_m$, from real galaxy
  clustering data sets.

    The construction of proper training samples
    would require running N-body or hydrodynamic simulations in a
    sufficiently large set of cosmological scenarios. As already
    noted, this task demands substantial computational resources and a
    dedicated work, which is beyond the scope of the current
    analysis. A forthcoming paper will be dedicated to this
    fundamental task. Here instead we rely on log-normal mock
    catalogues, which can reproduce the monopole of the redshift-space
    2PCF of real galaxy catalogues with reasonable accuracy and in a
    minimal amount of time. This will allow us to test our NN on
    training data sets with the desired format, in order to be ready
    for forthcoming analyses with more reliable data sets. Moreover,
    future developments of the presented method will involve
    estimating a higher number of cosmological parameters at the same
    time. The implemented NN is provided through a public Google
  Colab notebook\footnote{The notebook is available at:
\href{https://colab.research.google.com/drive/1zVxY9G5N9SNbsdjNHztVNpEhcTa6DBfu?usp=sharing}{Colab}.}.

The paper is organised as follows. In Section \ref{sec:method} we give
a general overview of the data analysis method we present in this
work. In Section \ref{sec:creat} we describe in detail the
characteristics of the catalogues used to train, validate and test the
NN. The specifics on the NN itself, together with its results on the
test set of mock catalogues are described in Section \ref{sec:NN}.
The application to the BOSS galaxy catalogue and the results this
leads to are presented in Section \ref{sec:application}.  In Section
\ref{sec:conclusion} we draw our conclusions and discuss about
possible future improvements. Finally, \ref{Appendix} provides details
on the algorithms used to construct the log-normal mock catalogues.


\section{The data analysis method}
\label{sec:method}
The data analysis method considered in this work exploits a supervised
machine learning approach. Our general goal is to extract cosmological
constraints from extra-galactic redshift surveys with properly trained
NNs. As a first application of this method, we focus the current
analysis on the 2PCF of BOSS galaxies, which is exploited to extract
constraints on $\Omega_m$.

The method consists of a few steps. Firstly, a fast enough process to
construct the data sets with which train, validate and test the NN is
needed. In this work we train the NN with a set of 2PCF mock
measurements obtained from log-normal catalogues. The construction of
these input data sets is described in detail in Section
\ref{sec:creat}. Specifically, we create several sets of mock
BOSS-like catalogues assuming as values for the cosmological
parameters the ones inferred from the Planck Cosmic Microwave
Background observations, except for $\Omega_m$, that assumes different
values in different mock catalogues, and $\Omega_\Lambda$ that has
been changed every time in order to have $\Omega_{tot}=\Omega_m+
\Omega_\Lambda+\Omega_{rad}=1$, where $\Omega_\Lambda$ and
$\Omega_{rad}$ are the dark energy and the radiation density
parameters, respectively.  Specifically, we fix the main parameters of
the $\Lambda$-cold dark matter ($\Lambda$CDM) model to the following
values: $\Omega_b\,h^2=0.02237$, $\Omega_c\,h^2=0.1200$,
$100\,\Theta_{MC}=1.04092$, $\tau=0.0544$,
$\textrm{ln}(10^{10}\,A_s)=3.044$, and $n_s=0.9649$ \citep[][Table 2,
  TT,TE+lowE+lensing]{Planck2018}.  Here $h$ indicates one hundredth
of the Hubble constant, $H_0$.  

The implemented NN performs a regression analysis, that is it can
assume every value within the range the algorithm has been trained
for, that in our case is $0.24<\Omega_m<0.38$. In particular, the
algorithm takes as input a 2PCF monopole measurement, and provides as
output a Gaussian probability distribution on $\Omega_m$, from which
we can extract the mean and standard deviation (see Section
\ref{sub:mocks} and \ref{Appendix}). Specifically, the data sets
  used for training, validating and testing the NN consist of
  different sets of 2PCF monopole measures, labelled by the value of
  $\Omega_m$ assumed to construct the mock catalogues these measures
  have been obtained from, and assumed also during the measure
  itself.

After the training and validation phases, we test the NN with a set of
input data sets constructed with random values of $\Omega_m$, that is,
different values from the ones considered in the training and
validation sets.  The structure of the NN, the characteristics of the
different sets of data it has been fed with, and how the training
process of the NN has been led are described in Section \ref{sec:NN}.

Finally, once the NN has proven to perform correctly on a test set of
2PCF measures from the BOSS log-normal mock catalogues, we exploit it
on the real 2PCF of the BOSS catalogue. To measure the latter sample
statistic, a cosmological model has to be assumed, which leads to
geometric distortions when the assumed cosmology is different to the
true one. To test the impact of these distortions on our final
outcomes, we repeat the analysis measuring the BOSS 2PCF assuming
different values of $\Omega_m$.  We find that our NN provides in
output statistically consistent results independently of the assumed
cosmology. The results of this analysis are described in detail in
Section \ref{sec:application}.

We provide below a summary of the main steps of the data analysis
method investigated in this work:

\begin{enumerate}

\item {\em Assume a cosmological model.}  In the current analysis, we
  assume a value of $\Omega_m$, and fix all the other main parameters of
  the $\Lambda$CDM cosmological model to the Planck values. A more extended
  analysis is planned for a forthcoming
  work.
  
\item {\em Measure the 2PCF of the real catalogue to be
  analysed and use it to estimate  the large-scale galaxy bias.}
  This is required to construct mock galaxy catalogues
  with the same clustering amplitude of the real data. Here we
  estimate the bias from the redshift-space monopole of the 2PCF at
  large scales (see Section \ref{sub:bias}).
  
\item {\em Construct a sufficiently large set of mock catalogues.}  In
  this work we consider log-normal mock catalogues, that can be
  obtained with fast enough algorithms. The measured 2PCFs of these catalogues
  will be used for the training, the validation and the test of the NN.

\item {\em Repeat all the above steps assuming different cosmological
  models}. Different cosmological models, characterised by different
  values of $\Omega_m$, are assumed to create different classes of mock
  catalogues and measure the 2PCF of BOSS galaxies. 
  
\item {\em Train and validate the NN.}

\item {\em Test the NN.}  This has to be done with data sets
  constructed considering cosmological models not used for the training and
  validation, to check whether the model can make reliable predictions
  also on previously unseen examples.

\item {\em Exploit the trained NN on real 2PCF measurements.} This is done
    feeding the trained machine learning model with the several measures of the 2PCF of the real catalogue, obtained assuming different cosmological models.
    The reason for this is to check whether the output
    of the NN is affected by geometric distortions.
\end{enumerate}


\section{Creation of the data set}
\label{sec:creat}

\subsection{The BOSS data set}
\label{sub:datasets}

The mock catalogues used for the training, validation and
test of our NN are constructed to reproduce the clustering of the BOSS galaxies. BOSS is part
of the Sloan Digital Sky Survey (SDSS), which is an imaging and
spectroscopic redshift survey that used a $2.5$m modified
Ritchey-Chr\'{e}tien altitude-azimuth optical telescope located at the
Apache Point Observatory in New Mexico \citep{Gunn2006}. The data we
have worked on are from the Data Release 12 (DR12), that is the final
data release of the third phase of the survey (SDSS-III)
\citep{Alam2015}.  The observations were performed from fall 2009 to
spring 2014, with a 1000-fiber spectrograph at a resolution
$R\approx2000$. The wavelength range goes from $360$ nm to $1000$ nm,
and the coverage of the survey is $9329$ square degrees.

The catalogue from BOSS DR12, used in this work, contains the
positions in observed coordinates (RA, Dec and redshift) of
$\num{1198004}$ galaxies up to redshift $z=0.8$.

Both the data and the random catalogues are created considering the
survey footprint, veto masks and systematics of the survey, as
e.g. fiber collisions and redshift failures. The method used to
construct the random catalogue from the BOSS spectroscopic
observations is detailed in \citet{Reid2016}.


\subsection{Two-point correlation function estimation}
\label{sub:2PCF}
We estimate the 2PCF monopole of the BOSS and mock galaxy
catalogues with the \citet{Landy1993} estimator:
\begin{equation}\label{landyszalay}
  \xi(s) = 1+\frac{N_{\rm RR}}{N_{\rm DD}}\frac{DD(s)}{RR(s)} -
  2\frac{N_{\rm RR}}{N_{\rm DR}}\frac{DR(s)}{RR(s)}\; ,
\end{equation}
where $DD(s)$, $RR(s)$ and $DR(s)$ are the number of galaxy-galaxy,
random-random and galaxy-random pairs, within given comoving
separation bins (i.e. in $s\pm\Delta s$), respectively, while $N_{\rm
  DD}=N_{D}(N_{D}-1)/2$, $N_{\rm RR}=N_{R}(N_{R}-1)/2$ and $N_{\rm
  DR}=N_{D}N_{R}$ are the corresponding total number of galaxy-galaxy,
random-random and galaxy-random pairs, being $N_{D}$ and $N_{R}$ the
number of objects in the real and random catalogues. The random catalogue is constructed with the same angular and redshift selection functions of the BOSS catalogue, but with a number of objects that is ten times bigger than the observed one, to minimise the impact of Poisson errors in random pair counts.


\subsection{Galaxy bias}
\label{sub:bias}

As introduced in the previous Sections, we train our NN to learn the
mapping between the 2PCF monopole shape, $\xi$, and the matter density
parameter, $\Omega_m$. Thus, we construct log-normal mock catalogues
assuming different values of $\Omega_m$. A detailed description of the algorithm used to construct these log-normal mocks is provided in the next Section \ref{sub:mocks}.

Firstly, we need to estimate the bias of the objects in the sample,
$b$. We consider a linear bias model, whose bias value is estimated
from the data. Specifically, when a new set of mock catalogues
(characterised by $\Omega_m=\Omega_{m,i}$) is constructed, a new
linear galaxy bias has to be estimated. The galaxy bias is assessed by
modelling the 2PCF of BOSS galaxies in the scale range
$30<s\,[h^{-1}\,\mbox{Mpc}]<50$, where $s$ is used here, instead of
$r$, to indicate separations in redshift space. We consider a Gaussian
likelihood function, assuming a Poissonian covariance
matrix which is sufficiently accurate for the
  purposes of this work.  We model the shape of the redshift-space
2PCF at large scale as follows \citep{Kaiser1987}:
\begin{equation}\label{model}
  \xi_{\rm gal}(s) = \left[(b\sigma_{8})^{2}+\frac{2}{3}f\sigma_{8} +
    \frac{1}{5}(f\sigma_{8})^{2}\right]\frac{1}{\sigma_{8}^{2}}\xi_m(r)\; ,
\end{equation}
where the matter 2PCF, $\xi_m(r)$, is obtained by Fourier
transforming the matter power spectrum modelled with the Code for
Anisotropies in the Microwave Background (\texttt{CAMB}) \citep{Lewis2000}.
The product between the linear growth rate and the matter power spectrum
normalisation parameter, $f\sigma_{8}$, is set by the cosmological model
assumed during the measuring of the 2PCF and the construction of the
theoretical model. The values of all redshift-dependent parameters have
been calculated using the mean redshift of the data catalogue, $z=0.481$.

We assume a uniform prior for $b\sigma_8$ between $0$ and $2.7$. The
posterior has been sampled with a Markov Chain Monte Carlo algorithm
with $10\,000$ steps and a burn-in period of $100$ steps. The linear
bias, $b$, is then derived by dividing the posterior median by
$\sigma_{8}$. The latter is estimated from the matter power spectrum
computed assuming Planck cosmology. We note that
  this method implies that only the shape of the 2PCF is actually used
  to constrain $\Omega_m$.


\subsection{Mock galaxy catalogues}
\label{sub:mocks}

As described in Section \ref{sec:method}, the data sets used for the
training, validation and test of the NN implemented in this work
consist of 2PCF mock measurements estimated from log-normal galaxy
catalogues.  Log-normal mock catalogues are generally used to create
the data set necessary for the covariance matrix estimate, in particular in
anisotropic clustering analyses \citep[see e.g.][]{lippich19}. In fact,
this technique allows us to construct density fields and, thus, galaxy
catalogues with the required characteristics in an extremely fast way,
especially if compared to N-body simulations, though the latter are more
reliable at higher-order statistics and in the fully nonlinear regime.

To construct the log-normal mock catalogues, we
  adopted the same strategy followed by \citet{beutler2011}. We
  provide here a brief explanation of the method, while a more
  detailed description is given in \ref{Appendix}. As already
  commented in Section \ref{sec:intro}, log-normal catalogues do not
  provide the optimal training sets for this machine learning method,
  but are used here only to test the algorithms with input data with
  the desired format.

The algorithm used in this work to generate these mock catalogues for
the machine learning training process takes as input one data
catalogue and one corresponding random catalogue. These are used to
define a grid with the same geometric structure of the data catalogue
and a visibility mask function which is constructed from the pixel
density of the random catalogue. A cosmological model has to be
assumed to compute the object comoving distances from the observed
redshifts and to model the matter power spectrum. The latter is used
to estimate the galaxy power spectrum, which is the logarithm of the
variance of the log-normal random field.

The density field is then sampled from this random
field. Specifically, once the algorithm has associated to all the grid
cells their density value, we can extract from each of them a certain
number of points that depend on the density of the catalogue and on
the visibility function. These points represent the galaxies of the
output mock catalogue.

Once the mocks are created, the same cosmological
  model is assumed also to measure the 2PCF.  We repeated the same
  process for all the test values of $\Omega_m$.  That is, for each
  $\Omega_m$ we created a set of log-normal mock catalogues and
  measured the 2PCF. This is different with respect to standard
  analyses, where a cosmology is assumed only once for the clustering
  measurements, and geometric distortions are included in the
  likelihood.

\begin{figure}[ht]
  \centering
  \includegraphics[width=0.49\textwidth]{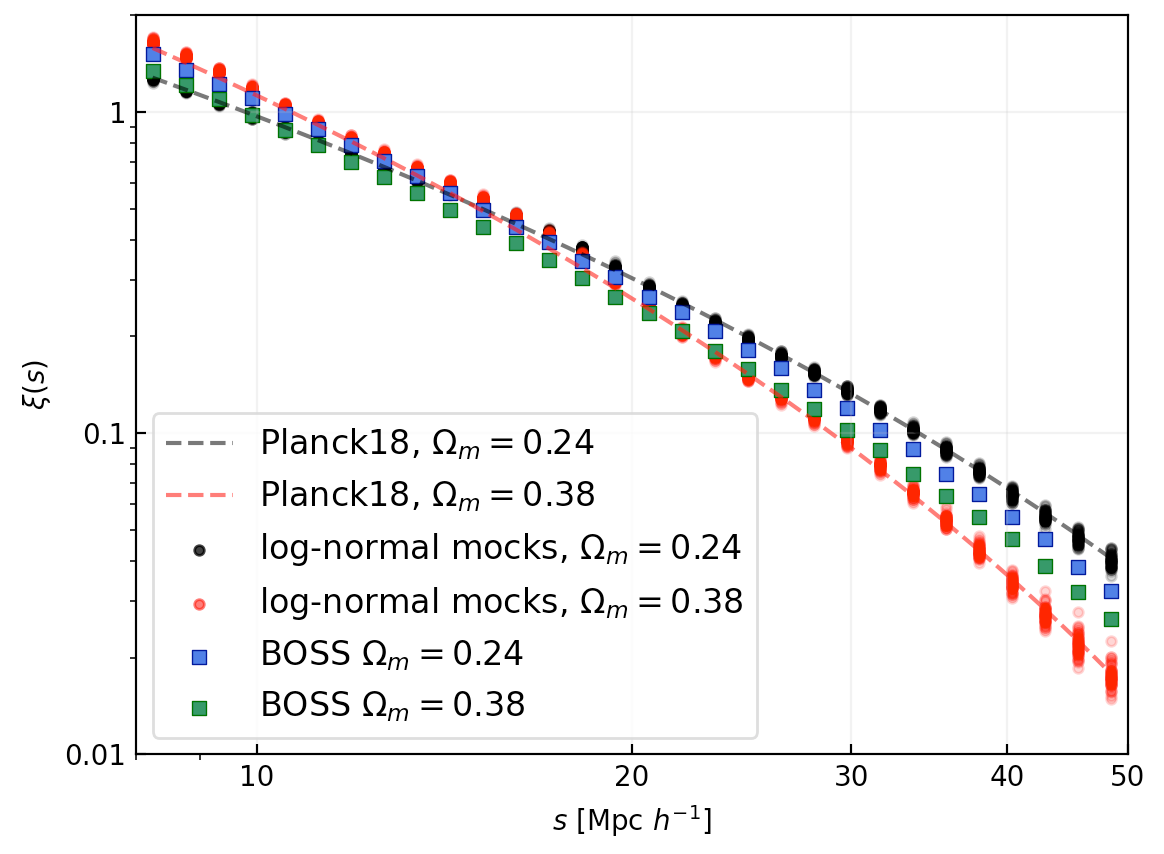}
  \caption{Different measures of the 2PCF monopole. The dots represent
    the measures obtained from $50$ log-normal mock catalogues
    constructed with $\Omega_m=0.24$ (black dots) and $\Omega_m=0.38$
    (red dots). The black and red dashed lines show the corresponding
    theoretical 2PCF models.  The blue and green squares show the real
    BOSS 2PCFs measured assuming the same cosmologies of the mock
    catalogues, that is $\Omega_m=0.24$ and $\Omega_m=0.38$,
    respectively, when converting redshifts into comoving
    coordinates.}
  \label{BOSS_vs_lognormal}
\end{figure}

As an illustrative example, Figure \ref{BOSS_vs_lognormal} shows
different measures of the 2PCF obtained in the scale range
$8<s\,[h^{-1}\,\mbox{Mpc}]<50$, in $30$ logarithmic bins of $s$, and
assuming the lowest and the highest values of $\Omega_m$ that have been
considered in this work, that are $\Omega_m=0.24$ and
$\Omega_m=0.38$.  The black and red dots show the measures
obtained from the corresponding two sets of $50$ log-normal mock
galaxy catalogues, while the dashed lines are the theoretical 2PCF models
assumed to construct them. As expected, the average 2PCFs of the two sets
of log-normal catalogues are fully consistent with the
corresponding theoretical predictions. Indeed, a mock log-normal
catalogue characterised by $\Omega_m=\Omega_{m,i}$
provides a training example for the NN describing how galaxies would
be distributed if $\Omega_{m,i}$ was the true value. Finally,
the blue and green squares show the 2PCFs of the real BOSS catalogue
obtained by assuming $\Omega_m=0.24$ and $\Omega_m=0.38$,
respectively, when converting galaxy redshifts into comoving
coordinates. The differences in the latter two measures are caused by
geometric distortions. As can be seen, neither of these two data sets
are consistent with the corresponding 2PCF theoretical models, that is
both $\Omega_m=0.24$ and $\Omega_m=0.38$ appear to be bad
guesses for the real value of $\Omega_m$.  As we will show in Section
\ref{sec:application}, the NN presented in this work is not
significantly affected by geometric distortions.

\section{The Artificial Neural Network}
\label{sec:NN}

\subsection{Architecture}

Regression machine learning models take as input a set of variables that can 
assume every value and provide as output a continuous value. In our case,
the input data set is a 2PCF measure, while the output is the predicted Gaussian
probability distribution of $\Omega_m$. Specifically, the regression model
we are about to describe has been trained with 2PCF measures in the scale range
$8-50 h^{-1}\,\mbox{Mpc}$, in $30$ logarithmic scale bins.

\begin{figure}[ht]
    \centering
    \includegraphics[width=0.49\textwidth]{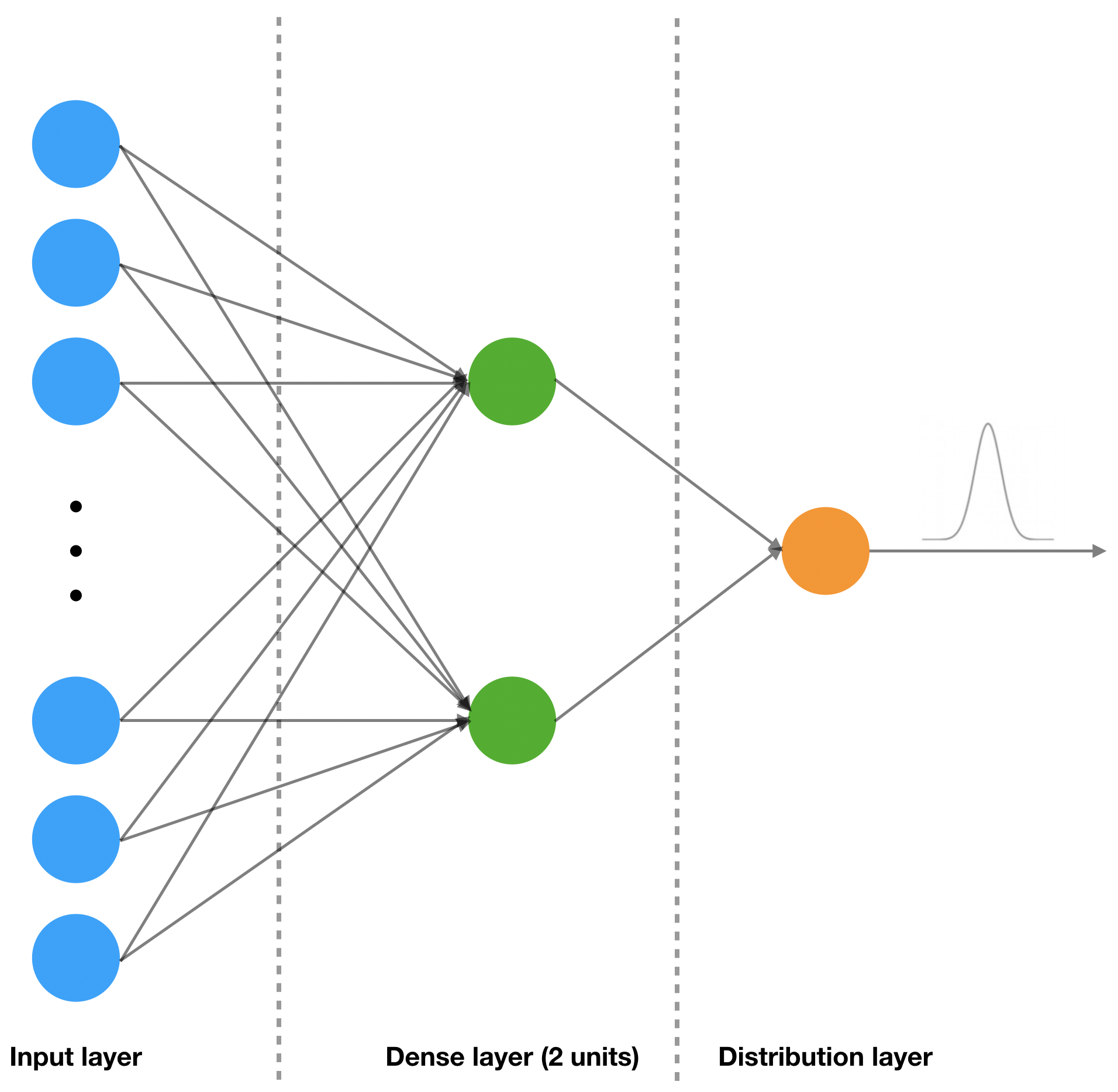}
    \caption{Schematic representation of the regression model
      considered in this work.  The input layer is represented by blue
      dots, the hidden dense layer by green dots, and the output layer
      by the orange dot.}
    \label{regr_model}
\end{figure}

The architecture of the implemented NN is schematically represented in
Figure \ref{regr_model}. It consists of the following layers:
\begin{itemize}
    \item Input layer that feeds the $30$ values of the measured 2PCF
      to the model;
    \item Dense layer\footnote{A layer is called dense if all its
    units are connected to all the units of the previous layer.} with
      2 units. Every unit of this layer performs
        the following linear operation on the input elements:
    \begin{equation}
        y=\sum_i w_i\cdot x_i+b\ \ ,
    \end{equation}
    where $y$ is the output of the unit, $x_i$ is the $i$-th element
    of the input, $w_i$ is its corresponding weight, and $b$ is the
    intercept of the linear operation. No nonlinear activation
    function has been used in this layer. This choice has been proven
    to be convenient a posteriori, providing excellent performance of
    the NN during the test phase (see Section \ref{sebsec:test}). We
    thus decided not to introduce any activation function to keep the
    architecture of the regression model as simple as possible;
    \item Distribution layer that uses the two outputs of the dense layer as mean 
    and standard deviation to parameterise a Gaussian distribution, which is given as output.
\end{itemize}

This architecture has been chosen because it is the simplest one we
tested which is able to provide accurate cosmological
constraints. Deeper models have been tried out, but no significant
differences were spotted in the output predictions.


\subsection{Training and validation}

The training and validation sets are constructed separately, and
consist of $2\,000$ and $800$ examples, respectively.  Regression
machine learning models work better if the range of possible outputs
is well represented in the training and validation sets. We construct
mock catalogues with $40$ different values of $\Omega_m$: $29$ from
$\Omega_m=0.24$ to $\Omega_m=0.38$ with $\Delta\Omega_m=0.005$ and
$11$ from $\Omega_m=0.2825$ to $\Omega_m=0.3325$, also separated by $\Delta\Omega_m=0.005$. The latter are added to improve
the density of inputs in the region that had proven to be the one
where the predictions were more likely, during the first attempts of
the NN training. The training and validation sets consist of $50$ and
$20$ 2PCF measures for each value of $\Omega_m$, respectively.  All
the mock catalogues used for the training and validation have the same
dimension of the BOSS 2PCF measures.

The loss function used for the training process is the following:
\begin{equation}
  J = -\sum_{i=1}^{N}\log\left[\frac{1}{\sqrt{2\pi}\sigma_i}\exp{-\frac{(l_i-\mu_i)^2}{2\sigma_i^2}}\right] \; ,
\end{equation}
where $N$ is the number of examples, $l_i$ is the label correspondent
to the $i$-th 2PCF used for the training or the validation (i.e. the
true value of $\Omega_m$ that has been used to create the $i$-th mock
catalogue and to measure its 2PCF), while $\sigma_i$ and $\mu_i$ are
the standard deviation and the mean of the Normal probability
distribution function that is given as output for the same $i$-th 2PCF
example.

During the training process, we apply the Adam optimisation
\citep[see][for a detailed description of its functioning and parameters]{Kingma2014} with the following three different steps:
\begin{itemize}
    \item 750 epochs with $\eta=0.002$\,,
    \item 150 epochs with $\eta=0.001$\,,
    \item 100 epochs with $\eta=0.0005$\,,
\end{itemize}
where $\eta$ indicates the learning rate. During the three steps, the values of the other two parameters of this optimisation algorithm are kept fixed to $\beta_{1}=0.9$ and $\beta_{2}=0.999$, and the
training set was not divided into batches. Variations in the
parameters of the optimization have been performed and did not lead to
significantly different outputs.

Gradually reducing the learning rate during the training helps the
model to find the correct global minimum of the loss function
\citep{Ntampaka2019}. The first epochs, having a higher learning rate,
lead the model towards the area surrounding the global minimum, while
the last ones, having a lower learning rate, and therefore being able
to take smaller and more precise steps in the parameter space, have
the task to lead the model towards the bottom of that minimum.  In
this model, minimizing the loss function means both to drive the mean
values of the output Gaussian distributions towards the values of the
labels (i.e. the values of $\Omega_m$ corresponding to the inputs),
and to reduce as much as possible the standard deviation of those
distributions.

During the training process, the interval of the labels, that is
$0.24\leq\Omega_m\leq0.38$, has been mapped into $[0,1]$ through the
following linear operation:
\begin{equation}
  \label{labelnormalization}
  L = \frac{l-0.24}{0.14}\; ,
\end{equation}
where $l$ is the original label and $L$ is the one that belongs to the
$[0,1]$ interval.  Once the mean, $\mu_{L}$, and the standard
deviation, $\sigma_{L}$, of the predictions are obtained in
output, we convert them back to match the original interval of the
labels.
The standard deviation is then computed as follows:
\begin{equation}
  \frac{d\mu_{L}}{d\sigma_{L}} = \frac{d\mu_l}{d\sigma_l}\; ,
\end{equation}
where $\sigma$ is the standard deviation of $\mu$, so that the
uncertainty on $\mu$ is given by $0.14\,\sigma_{L}$.


\subsection{Test}
\label{sebsec:test}

The test set consists of $5$ different 2PCF measures obtained from
log-normal mock catalogues with $16$ different randomly generated
values of $\Omega_m$. Therefore, we have a total of $80$ measures.
The $\Omega_m$ predictions and uncertainties are estimated as the
mean and standard deviation of the Gaussian distributions obtained in output
\citep{Matthies2007, Der2009, Kendall2017, Russell2019}.  Our model is thus able to associate to every point of the input space an uncertainty on the output that depends on the intrinsic scatter of the 2PCF measures.

\begin{figure}[ht]
  \centering
  \includegraphics[width=0.49\textwidth]{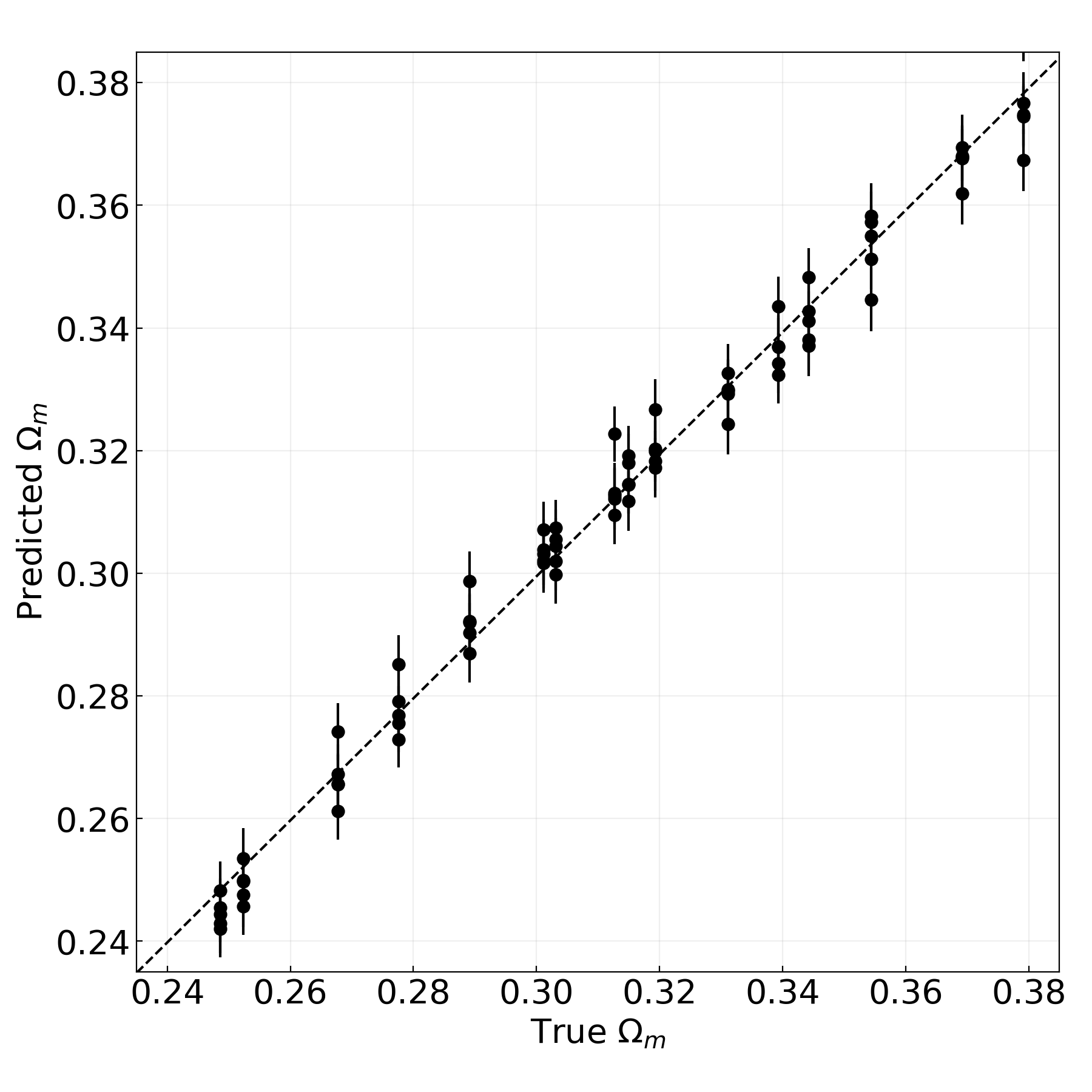}
  \caption{Values of $\Omega_m$ and related uncertainties predicted by
    our NN, compared with the best-fit linear model (dashed line). The
    best-fit slope is $0.995\pm0.014$, while the best-fit
    normalisation is $0.001\pm0.005$. The error bars in this plot represent
    the standard deviations of the distributions for $\Omega_m$ obtained as 
    output of the NN}
  \label{regr_res}
\end{figure}

Figure \ref{regr_res} shows the predictions on the test set, compared
to the true values of $\Omega_m$ the mocks in the test were
constructed with. As can be seen, the implemented NN is able to
provide reliable predictions when fed with measures from mock
catalogues characterised by $\Omega_m$ values that were not used
during the training and validation phases. In fact, fitting this data
set with a linear model:
\begin{equation}
  \Omega_{m}^{\rm pred} = \alpha\cdot\Omega_{m}^{\rm true}+\beta\; ,
\end{equation}
we get $\alpha=0.995\pm0.014$ and $\beta=0.001\pm0.005$, which are
consistent with the slope and the intercept of the bisector of the
quadrant. Figure \ref{regr_res} also shows that
  the estimates near the limits of the training range at
  $\Omega_m=0.24$ and $0.38$ are not biased, as the predictions on the
  test set, also for the most external examples, are consistent with
  the bisector of the quadrant.


\section{Application to BOSS data}
\label{sec:application}

After training, validating and testing the NN, we can finally apply it
to the real 2PCF of the BOSS galaxy catalogue.  To test the impact of
geometric distortions caused by the assumption of a particular
cosmology during the measure, we apply the NN on $40$ measures of the
2PCF of BOSS galaxies obtained with different assumptions on the value
of $\Omega_m$ when converting from observed to comoving coordinates.

\begin{figure}[ht]
  \centering
  \includegraphics[width=0.49\textwidth]{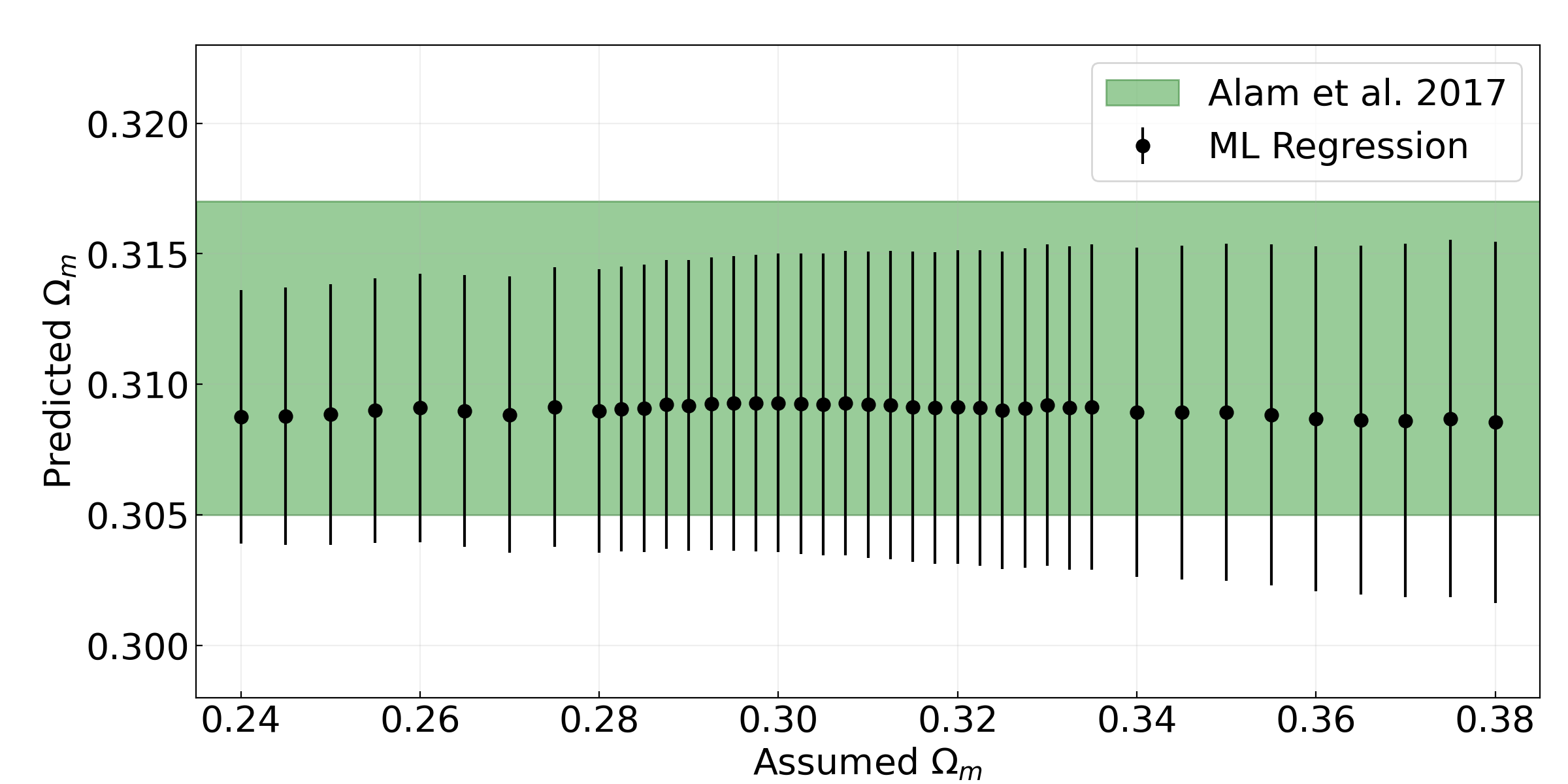}
  \caption{Machine learning predictions (black dots and bars) on
    $\Omega_m$ from the measured 2PCF of BOSS galaxies, as a function
    of the $\Omega_m$ assumed in the measurements. The predictions are
    compared to the $\Omega_m$ constraints provided by
    \citet{Alam2017}, here represented by the green stripe.}
  \label{BOSS_res}
\end{figure}

Figure \ref{BOSS_res} shows the results of this analysis, compared
with the $\Omega_m$ constraints provided by \citet{Alam2017}.
The predictions of the NN have been fitted with a linear model:
\begin{equation}
  \Omega_m^{\rm pred} = \alpha\cdot\Omega_{m}^{\rm ass}+\beta\; ,
\end{equation}
where $\Omega_m^{\rm pred}$ is the prediction of the regression model, while
$\Omega_{m}^{\rm ass}$ is the value assumed to measure the 2PCF.  The best
fit-values of the parameters we get are $\alpha=-0.001\pm0.025$ and
$\beta=0.309\pm0.008$. In particular, the slope is consistent with zero,
that is all the $\Omega_m$ predictions are consistent, within the
uncertainties, independently of the value assumed in the measurement.
This demonstrates that the NN is indeed able to make robust predictions
over observed data, without being biased by geometric distortions. 

Our final $\Omega_m$ constraint is thus estimated from the best-fit normalization, $\beta$, that is  
\begin{equation}
  \label{final_result_regression}
  \Omega_m = 0.309\pm0.008\; .
\end{equation}

This result is consistent with the one obtained by \citet{Alam2017},
which is $\Omega_m=0.311\pm0.006$.


\section{Conclusions}
\label{sec:conclusion}

In this work we investigated a supervised machine learning data
analysis method aimed at constraining $\Omega_m$ from observed galaxy
catalogues. Specifically, we implemented a regression NN that has been
trained and validated with mock measurements of the 2PCFs of BOSS
galaxies. Such measures are used as a convenient summary statistics of
the large-scale structure of the Universe. The goal of this work was
to infer cosmological constraints without relying on any analytic 2PCF
model to construct the likelihood function.

To train and validate our NN, we use $2\,800$ 2PCF examples,
constructed with $40$ different values of $\Omega_m$, in
$0.24\leq\Omega_m\leq0.38$. The trained NN has been finally applied to
the real 2PCF monopole of the BOSS galaxy catalogue. We get
$\Omega_m=0.309\pm0.008$, which is in good agreement
with the value found by \citet{Alam2017}.

This work confirms that NNs can be powerful tools also for
cosmological inference analyses, complementing
  the more standard analyses that make use of analytical likelihoods.
One obvious improvement of the presented work would be to consider
more accurate mock catalogues than the log-normal ones, for the
training, validation and test phases.  In particular, N-body or
hydrodynamic simulations would be required to exploit higher-order
statistics, as in particular the three-point correlation function, as
features to feed the model with.

The higher the number of reliable features is, the more accurate the
predictions of the NN will be. Furthermore, multi-labelled regression
models can be used to make predictions on multiple cosmological
parameters at the same time. To do that, however, bigger data sets are
required, in order to have a proper mapping of the input space,
characterised by the different values each label can have. The
analysis presented in this work should also be extended to larger
scales, though in this case more reliable mock catalogues are required
not to introduce biases in the training, in particular at the Baryon
Acoustic Oscillations scales. Finally, a similar analysis as the one
performed in this work could be done using the density map of the
catalogue, or directly the observed coordinates of the galaxies. This
approach would be less affected by the adopted data compression
methods considered. On the other hand, it would have a significantly
larger computational cost for the training, validation and test
phases, which should be estimated with a dedicated feasibility
study. All the possible improvements described above will be
investigated in forthcoming papers.


\section*{Acknowledgements}
We thank the anonymous referee for her/his comments
  that helped to improve the presentation of our results. We
acknowledge the use of computational resources from the parallel
computing cluster of the Open Physics Hub
(https://site.unibo.it/openphysicshub/en) at the Physics and Astronomy
Department in Bologna. FM and LM acknowledge the grants ASI
n.I/023/12/0 and ASI n.2018-23-HH.0. LM acknowledges support from the
grant PRIN-MIUR 2017 WSCC32.

{\em Software}: \texttt{Numpy} \citep{harris2020}; \texttt{Matplotlib} \citep{Hunter2007};
\texttt{SciPy} \citep{virtanen20}; \texttt{CosmoBolognaLib} \citep{Marulli2016};
\texttt{CAMB} \citep{Lewis2000}; \texttt{FFTLog} \citep{Hamilton2000};
\texttt{TensorFlow} \citep{tensorflow2015}; \texttt{Keras} \citep{chollet2015}.


\appendix

\section{Log-normal mock catalogues}
\label{Appendix}

In the following, we describe the algorithm used
  in this work to construct log-normal mock catalogues. We followed
  the same strategy as in \citet[][see their Appendix
    B]{beutler2011}. Let us define a random field in a given volume
as a field whose value at the position $\Vec{r}$ is a random variable
\citep{Peebles1993, Xavier2016}. One example is the Gaussian random
field, $N(\Vec{r})$. In this case the one-point probability density
function (PDF) is a Gaussian distribution, fully characterised by the
mean, $\mu$, and the variance, $\sigma^{2}$.  If $n$ positions are
considered, instead of just one, the PDF is the multivariate Gaussian
distribution:
\begin{equation}
  \label{multivariategaussian}
  f_{n}(\Vec{N}) = \frac{1}{(2\pi)^{n/2}|\boldsymbol{M}|^{1/2}}
  \exp\left[-\frac{1}{2}\sum_{i,j}\boldsymbol{M}_{ij}^{-1}N_{i}N_{j}\right]\;
  ,
\end{equation}
where $N_{i}=N(\Vec{r}_{i})$, and $\boldsymbol{M}$ is the covariance
matrix, the elements of which are defined as follows:
\begin{equation}
  \label{covariance}
  \boldsymbol{M}_{ij} = \langle(N_{i}-\mu)(N_{j}-\mu)\rangle\; .
\end{equation}
The PDF of the primordial matter density contrast at a specific
position, $\delta(\mathbf{x})$, can be approximated as a Gaussian
distribution, with null mean and the correlation function as
variance. The same definitions can be used in Fourier space as
well. In this case the variance is the Fourier transform of the 2PCF,
that is the power spectrum, $P(k)$. In more general cases, the
Gaussian random field is only an approximation of the real random
field that may present features, such as significant skewness and
heavy tails \citep{Xavier2016}.

\citet{Coles1987} showed how to construct non-Gaussian fields through
nonlinear transformations of a Gaussian field. One example is the
log-normal random field \citep{Coles1991}, which can be obtained
through the following transformation:
\begin{equation}
  \label{lognormaltransformation}
  L(\Vec{r}) = \exp[N(\Vec{r})]\; .
\end{equation}
The log-normal transformation results in the following one-point PDF:
\begin{equation}
  f_{1}(L) = \frac{1}{\sqrt{2\pi\sigma^{2}}}
  \exp\left[-\frac{(\log(L)-\mu)^{2}}{2\sigma^{2}}\right]\frac{dL}{L}
  \; ,
\end{equation}
where $\mu$ and $\sigma^{2}$ are the mean and the variance of the
underlying Gaussian field $N$, respectively.

The multivariate version for the log-normal random field is defined as
follows:
\begin{equation}
  f_{n}(\Vec{L}) = \frac{1}{(2\pi)^{n/2}|\boldsymbol{M}|^{1/2}}
  \exp\left[-\frac{1}{2}
    \sum_{i,j}\boldsymbol{M}_{ij}^{-1}\log(L_{i})\log(L_{j})\right]\prod_{i=1}^{n}\frac{1}{L_{i}}
  \; ,
\end{equation}
where $\boldsymbol{M}$ is the covariance matrix of the N-values.

To construct the log-normal mock catalogues used
  to train, validate and test our NN, we start from a power spectrum
  template, assuming a value for the bias. After creating a grid
  according to this power spectrum, a Fourier transform is performed
  to obtain the 2PCF, which is used to calculate the function
  $\log[1+\xi(r)]$.  We then revert to Fourier-space obtaining a
  modified power spectrum, $P_{ln}(k)$, and assign to each point of
  the grid a value of the Fourier amplitude, $\delta(k)$, sampled from
  a Normal distribution that has $P_{ln}(k)$ as standard deviation. A
  new Fourier transform is performed to obtain a density field
  $\delta(x)$ from which we sample the galaxy distribution. In our
  calculations we used a grid size of $10\,\mbox{Mpc}\,h^{-1}$, while
  the minimum and the maximum values of $k$ are
  $0.0001\,h\,\mbox{Mpc}^{-1}$ and $100\,h\,\mbox{Mpc}^{-1}$,
  respectively.

\bibliographystyle{elsarticle-harv} \bibliography{main.bib}

\end{document}